\documentclass[]{spie}  
\usepackage[]{graphicx}

\newcommand{\degrees}[0]{\ensuremath{^{o}}}

\newcommand{\loopg}[0]{\ensuremath{\mathscr{L}}}
\newcommand{\ohms}[0]{\ensuremath{\Omega}}

\newcommand{\revOne}[1]{#1}
\newcommand{\revOneOmit}[1]{}
\newcommand{\revTwo}[1]{#1}

\newcommand{\revTwoOmit}[1]{}

\newcommand{\acomment}[1]{}

\usepackage{SIunits}
\usepackage{subfig}
\usepackage{mathrsfs}
\usepackage{url}
\usepackage[usenames,dvipsnames]{color}

\title{The performance of the bolometer array and readout system during the 2012/2013 flight of the E and B experiment (EBEX)} 

\author{Kevin MacDermid\supit{a}, Asad M. Aboobaker\supit{c}, Peter Ade\supit{d}, Fran\c{c}ois Aubin\supit{c},
Carlo Baccigalupi\supit{e}, Kevin Bandura\supit{a}, Chaoyun Bao\supit{c}, Julian Borrill\supit{f}, Daniel Chapman\supit{g}, Joy Didier\supit{g}, Matt Dobbs\supit{a}\supit{s}, Julien Grain\supit{h}, Will Grainger\supit{d},
Shaul Hanany\supit{c}, Kyle Helson\supit{k}, Seth Hillbrand\supit{d}, Gene Hilton\supit{r}, Hannes Hubmayr\supit{r},
Kent Irwin\supit{r}, Bradley Johnson\supit{d}, Andrew Jaffe\supit{i}, Terry Jones\supit{c}, Ted Kisner\supit{i},
Jeff Klein\supit{c}, Andrei Korotkov\supit{k}, Adrian Lee\supit{b}, Lorne Levinson\supit{l}, Michele Limon\supit{g}, Amber Miller\supit{g}, Michael Milligan\supit{c}, Enzo Pascale\supit{d}, Kate Raach\supit{c},
Britt Reichborn-Kjennerud\supit{g}, Carl Reintsema\supit{r}, Ilan Sagiv\supit{l}, Graeme Smecher\supit{a},
Radek Stompor\supit{o}, Matthieu Tristram\supit{q}, Greg Tucker\supit{k},  Ben Westbrook\supit{b}, Kyle Zilic\supit{c}
\skiplinehalf
\supit{a}McGill University, Montr\'eal, Quebec, H3A 2T8, Canada;\\
\supit{b}University of California, Berkeley, Berkeley, CA 94720;\\
\supit{c}University of Minnesota School of Physics and Astronomy, Minneapolis, MN 55455;\\
\supit{d}Rutherford Appleton Lab, Harwell Oxford, OX11 0QX;\\
\supit{e}Scuola Internazionale Superiore di Studi Avanzati, Trieste 34014, Italy;\\
\supit{f}Space Sciences Laboratory, University of California, Berkeley, California, U.S.A. ; \\
\supit{g}Columbia University, New York, NY 10027;\\
\supit{h}Institut d'Astrophysique Spatiale, Universite Paris-Sud, Orsay, 91405, France;\\
\supit{i}Lawrence Berkeley National Laboratory, Berkeley, California, U.S.A.;\\
\supit{k}Brown University, Providence, RI 02912;\\
\supit{l}Weizmann Institute of Science, Rehovot 76100, Israel;\\
\supit{m}California Institute of Technology, Pasadena, CA 91125;\\
\supit{o}Laboratoire Astroparticule et Cosmologie (APC), Universit\'e Paris Diderot, 75205, France;\\
\supit{p}University of California, Berkeley, Space Sciences Lab, Berkeley, CA 94720;\\
\supit{q}Laboratoire de l'Accelerateur Lineaire, Universite Paris-Sud, Orsay, 91405, France;\\
\supit{r}National Institute of Standards and Technology, MD 2089;\\
\supit{s}Canadian Institute for Advanced Research, CIFAR Program in Cosmology and Gravity, Toronto, ON, M5G1Z8, Canada
}

\begin{document} 
\maketitle 

\begin{abstract}
EBEX is a balloon-borne telescope designed to measure the polarization of the cosmic microwave background radiation. 
During its eleven day science flight in the Austral Summer of 2012, it operated 955 spider-web transition edge 
sensor (TES) bolometers separated into bands at 150, 250 and 410~GHz. This is the first time that an array of TES bolometers
has been used on a balloon platform to conduct science observations. Polarization sensitivity was provided by a wire grid 
and continuously rotating half-wave plate. The balloon implementation of the bolometer array and readout electronics 
presented unique development requirements. Here we present an outline of the readout system, the remote tuning of 
the bolometers and Superconducting QUantum Interference Device (SQUID) amplifiers, and preliminary current noise 
of the bolometer array and readout system. 
\end{abstract}


\section{Introduction}
The E and B experiment (EBEX) is a balloon-borne telescope designed to measure the polarization of the cosmic microwave
 background radiation (CMB) from degree to 10 arcminute scales. Science goals include
measurements of CMB E and B-modes and the polarization of galactic dust. It was launched from William's Field, 
Antarctica on December 29th, 2012 for its science flight. The payload circumnavigated the continent at an altitude of 
approximately 35~km until the flight was terminated on January 23rd, 2013. Science data was collected for the first 
eleven days of this flight, until the liquid cryogens required to cool the SQUIDs and detectors were expended. 

EBEX is the first payload to implement transition edge sensor bolometers (TES) during its engineering flight in 2009, and then 
a kilo-pixel array for the 2012 science flight. The experiment had three frequency
bands centered at 150, 250, and 410~GHz. The detectors were read out with SQUID-based, digital frequency 
domain multiplexed (DfMUX) readout electronics. This readout scheme was first implemented on EBEX, used on its engineering flight in 2009~\cite{Aubin2010}, and later adopted by several ground based 
experiments such as POLARBEAR~\cite{Polarbear} and SPTpol~\cite{Sptpol}. One of the key motivations for implementing 
DfMUX on EBEX was that
it was designed to consume about 10 times lower power than its analog predecessor~\cite{Dobbs2012_readout}. The total 
power consumption for a system that was able to supply readout for up to 1792 channels just under 600~W. In addition,
 the readout electronics provided embedded processors able to perform the tuning of the SQUIDs and bolometers 
 autonomously and in parallel, 
thereby reducing the computing requirements of the EBEX flight computers. In Section~\ref{sec:bolometerarray} 
and~\ref{sec:readoutsystem} we discuss the implementation of 
the bolometer array and readout system in more detail, and in Section~\ref{sec:flight performance} 
we give in-flight performance, including the 
SQUID and bolometer tuning procedures, yields, and current noise.

\section{Bolometer Array}
\label{sec:bolometerarray}

\revTwo{The EBEX optical design includes two focal planes, each consisting of seven bolometer wafers~\cite{Britt2010}~\cite{Milligan2011}. These focal planes are cooled to 260~mK by a 
 combination of liquid nitrogen and helium, which cool the receiver to 4~K, and two closed cycle
 helium-sorption refrigerators \cite{Sagiv2011}.}
 
Each bolometer wafer includes 140 individual spider-web
transition edge sensor (TES) bolometer pixels \cite{Irwin2005}.  Due to readout channel limitations we could only
readout a maximum of 128 bolometers per wafer. In practice a smaller number were wired. The lower yield was due to
fabrication loss and pre-flight selection based on bolometer characterization measurements. For flight, there were 679,
377, and 134 wired channels at 150, 250 and 410~GHz, respectively. Of these 1190 channels, 1105 were open to light, the remaining 85 were a combination of dark and resistor channels included to act as test channels for the readout electronics.

\begin{figure}[htbp]
\centering
\subfloat[]{\includegraphics[width=4.15 cm]{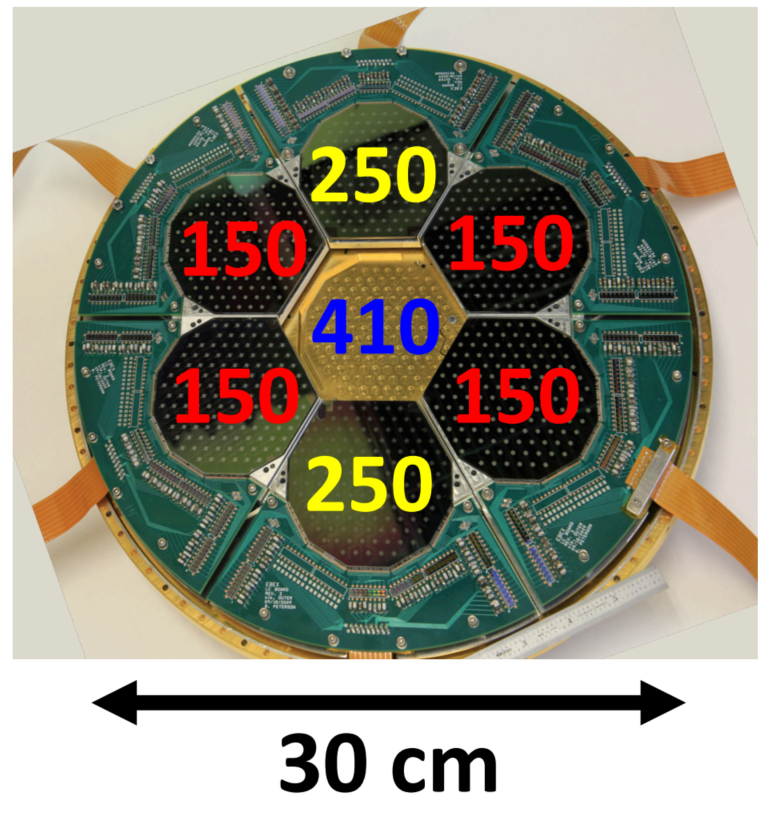}}
\subfloat[]{\includegraphics[width=3.55 cm]{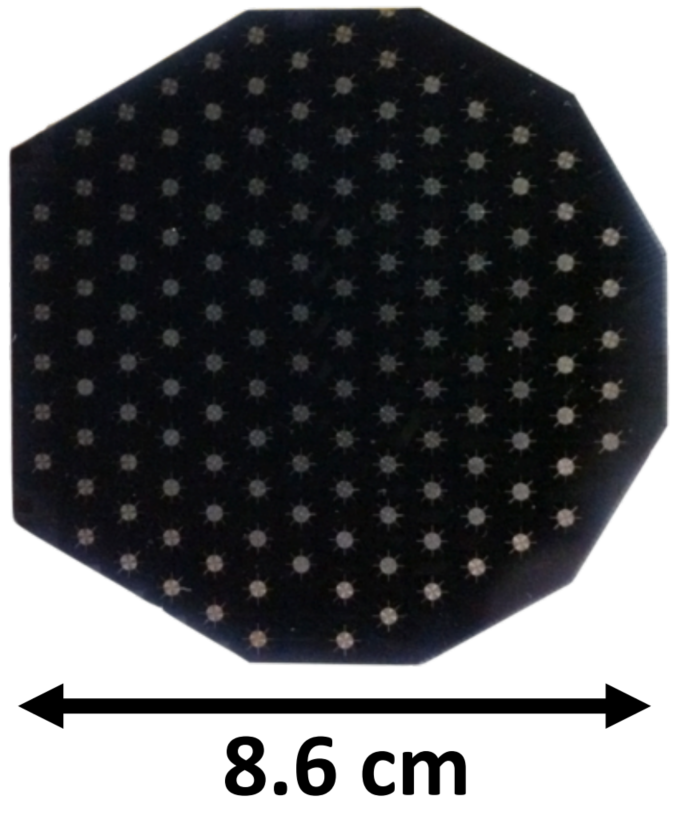}}
\subfloat[]{\includegraphics[width=4 cm]{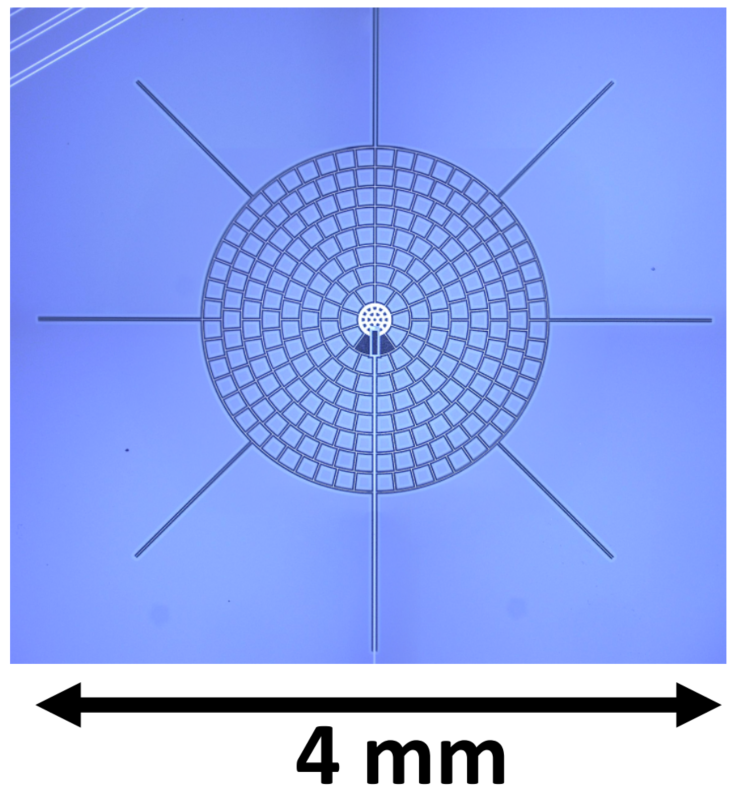}}
\subfloat[]{\includegraphics[width=4 cm]{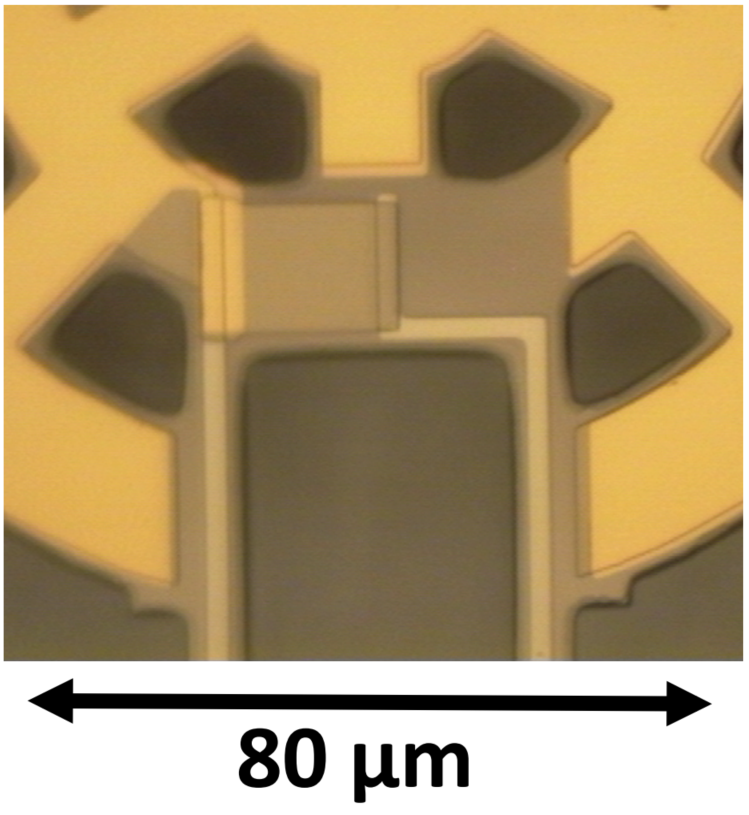}}
\caption{(a) A photograph of one of the EBEX focal planes with the frequency bands labelled. Six of the seven wafers are shown, the 410~GHz had not yet been installed. 
      (b) A photograph of a bolometer wafer. There are 140 bolometer channels on each wafer, visible as light colored circles.
      (c) A microscope image of a spider-web bolometer. The light (cream) shaded area is the silicon nitride layer on top of the wafer. The dark shading are trenches in silicon nitride in which a narrower silicon nitride web is suspended. The lines extending at 45\degrees{} increments are the trenches that surround eight silicon nitride legs that suspend the absorber and connect it to the wafer. The TES is at the bottom of the light section at the center of absorber. This is a gold thermal ballast designed to slow the thermal response of the TES.
      (d) A microscope image of the TES. It is a aluminium titanium sandwich where the thickness of the aluminium is tuned to set the critical temperature.}
\label{fig:tes}
\end{figure}

Figure~\ref{fig:tes}(a) is a photograph of one of the EBEX focal planes during commissioning. Each wafer position is 
labelled with its frequency band. Note that the 410~GHz wafer has not yet been installed in this photograph. 
Plate (b) is a photograph of a bolometer wafer, the 140 bolometers are visible as the lighter dots on the surface. 
Plate (c) is a microscope image of a bolometer pixel. Each bolometer consists of a gold-plated silicon nitride 
absorber lithographed into the shape of a
spider web. The eight legs extending radially at 45\degrees{} increments and the electrical leads 
form the weak thermal link between the absorber and the bath. The TES is at the center of the absorber, it is 
thermally connected to a gold structure to slow its thermal response. Plate (d) is a microscope image of the TES. 
It is a bilayer of aluminium and titanium where the thickness of the aluminium was tuned to set the critical temperature 
via the proximity effect \cite{Westbrook2012}.

The bolometers are not inherently polarization sensitive but are preceded by a continuously 
rotating broad-band half-wave plate and polarizing beam splitter, which splits
the incoming light, sending one linear polarization to each focal plane. The spectral band of each bolometer wafer is
set by low-pass metal mesh filters and waveguides for each pixel as high pass filters. 

\section{Readout System}
\label{sec:readoutsystem}

EBEX uses an array of over a thousand TES bolometers cooled to 260~mK. One challenge introduced by this pixel count is
the thermal loading on both the 4~K and 250~mK stages due to the readout wiring. This is addressed by multiplexing the 
detectors in the frequency domain which allows a number of bolometer channels to share a single pair of wires. During the 2012/2013 flight, we used a multiplexing factor of 16, the highest achieved at the time 
with this readout scheme. 

\begin{figure}[htbp]
\begin{center}
\includegraphics[width=12cm]{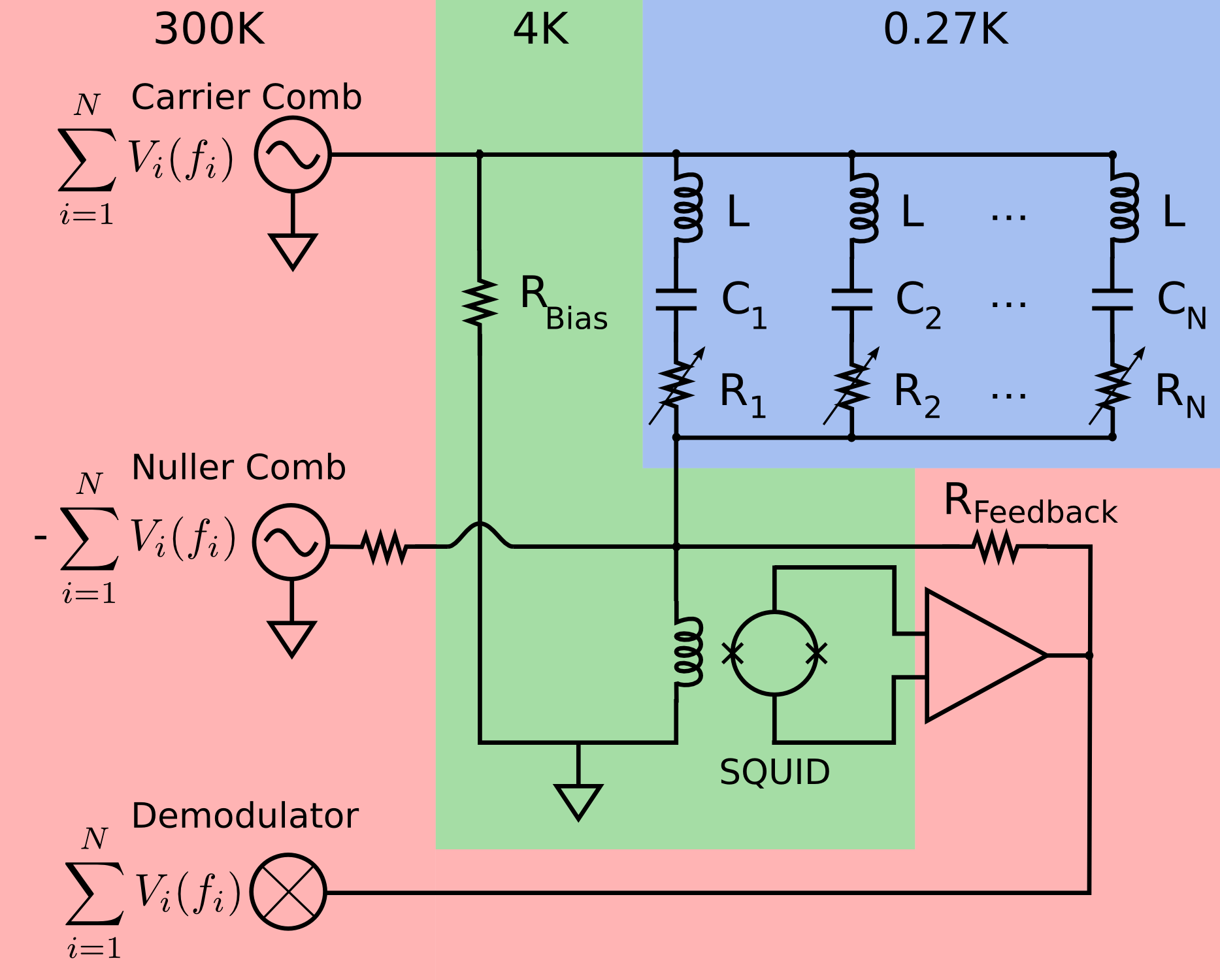} 
\caption{
A conceptual diagram of the frequency domain multiplexed readout circuit. The bolometers are the variable resistors in
the upper right, each forming part of an RLC circuit. The ``Carrier Comb'' in the upper left contains a carrier sinusoid
near the resonant frequency of each of these filters. These resonant frequencies are from 400 to 1200~kHz. These sinusoids provide the electrical voltage bias for the bolometers. The
``Nulling Comb'' eliminates the carriers at the SQUID coil, to avoid loading the SQUID unnecessarily. Finally, the
``Demodulator'' separates the signal for each channel by multiplying by a sinusoid at each of the frequencies present in
the carrier comb, low pass filtering and downsampling the result. The carrier, nuller and demodulator operate at 25~MHz, the final downsampled data rate is 190.73~Hz. The temperature of each section is indicated by its color.
}
\label{fig:cryogenic_circuit} 
\end{center}
\end{figure}

Figure~\ref{fig:cryogenic_circuit} is a conceptual diagram of the readout system. Each bolometer, shown as a variable
resistor in the upper-right of Figure~\ref{fig:cryogenic_circuit}, forms part of an inductive-capacitive-resistive (LCR)
filter. The voltage bias for each channel is supplied by a sinusoid with a frequency near the resonance of its filter. This voltage bias ($V_\mathrm{bias}$) causes the bolometer to exhibit negative electro-thermal feedback, improving its linearity, dynamic range and time constant. The carrier signals are synthesized and digitally summed by the warm readout electronics, creating the ``carrier
comb'' shown in the upper left of Figure~\ref{fig:cryogenic_circuit}. The resistance of each bolometer changes in
response to the incident sky signal, amplitude modulating the carrier such that sky signals appear in sidebands of the
carrier frequency~\cite{Smecher2010}.

The signals from all 16 bolometers on the same multiplexed module are summed at the SQUID input coil.  In addition, a
nuller comb is injected at the SQUID input coil which contains tones at the same frequency as those in the carrier comb,
but 180\degrees{} out of phase. This nuller comb serves to remove the carriers without affecting the sky signals, which
drastically reduces the dynamic range requirements of the SQUID by keeping the current through the input coil low.

The nulled signal is amplified by the SQUID and transmitted to the warm readout electronics. These consist of the SQUID
control board, and the digital frequency multiplexing (DfMUX) boards \cite{Dobbs2008}. The SQUID control board houses the amplifier and
SQUID feedback resistor ($R_\mathrm{feedback}$) shown in the lower right of Figure \ref{fig:cryogenic_circuit}. The
DfMUX board includes a Xilinx Virtex-4 Field-Programmable Gate Array (FPGA), eight Digital-to-Analog converters (DACs),
and four demodulator Analog-to-Digital (ADC). Each DfMUX produces the carrier and nuller combs, demodulates, and
channelizes all the bolometers connected to on pair of wires, typically called a ``multiplexed module.''

\newpage
\section{Flight Performance}
\label{sec:flight performance}
This section discusses the performance of the bolometers and readout system during the 2012/2013 science flight, including both the tuning procedure to operate the array and the SQUID and bolometer noise.

\subsection{Tuning}
\label{sec:tuning}

\revOne{Tuning of the SQUID and bolometer arrays is performed during each cycle of the closed-cycle helium sorption refrigerators that cool the focal plane and optics. Tuning SQUID and bolometer arrays on a balloon platform presented several challenges relative to ground-based implementations:}
%

\begin{itemize}
\item{\underline{Shorter observation time} - For EBEX, the sorption refrigerators were cycled every 2-3 days.
        Maintaining high observing efficiency requires that all service tasks be completed within few hours.  Therefore,
        we developed a procedure to interleave the tuning with the fridge cycle, taking advantage of waiting periods
        already present in the cycle due to the long thermal time constants involved in the boiling and condensation of
    the $^{3}$He and $^{4}$He in the sorption fridges.}

\item{\underline{Limited telemetry and commanding bandwidth} - Sending commands to the balloon was limited to about 150 bit/s
        \cite{Milligan2011}. This motivated the development of new commanding software that minimized the information
        transferred to the balloon during tuning requests while maintaining fine grained control over the tuning
        involved. This was achieved by storing the hardware connections between readout elements (such as bolometers
        and SQUIDs), as well as the tuning parameters in an sqlite3 database. This database was constructed
        using pre-flight tests and stored to the solid state disk on each of the flight computers.
        During flight we were able to tune specific sections of flight hardware, such as all SQUIDs wired to a
        particular readout crate, using custom commands that included SQL syntax, for flexibility and robustness \cite{sqlite}.}

\item{\underline{Limited computing resources} - The EBEX flight control computers (FCCs) were required to be low power,
        well tested in space-like conditions, and to integrate with other on-board systems~\cite{Milligan2011}.  Once
        the computer technology was selected, early in the development of EBEX, it was not possible to continually
        upgrade the flight computers. During the commissioning of the experiment we found that having the FCCs manage the
        entire tuning of the array places on it heavy computing and communication load. We therefore implemented new 
        firmware on the embedded processors of the DfMUX boards such that they would execute tuning tasks instead
        of the FCCs. When tuning occurred, either during a sorption refrigerator cycle or one initiated by ground operators, a 
        single command from the FCCs to any DfMUx board would trigger the board to cycle through the entire tuning process 
        autonomously and only communicate the end results to the FCCs.   }

\end{itemize}

An outline of the tuning process and performance are presented in sections \ref{sec:squid tuning} and \ref{sec:bolometer
tuning}. Further details of the EBEX tuning system are available in MacDermid, 2014 \cite{MacDermid2014}. 

\subsubsection{SQUID Tuning}
\label{sec:squid tuning}
\revOne{This section describes the SQUID tuning process. It begins with the goals of SQUID tuning, then desribes the algorithm used, and concludes with
the SQUID tuning yield during the flight.
}

\begin{figure}[htbp]
\begin{center}
    \includegraphics[width=5in]{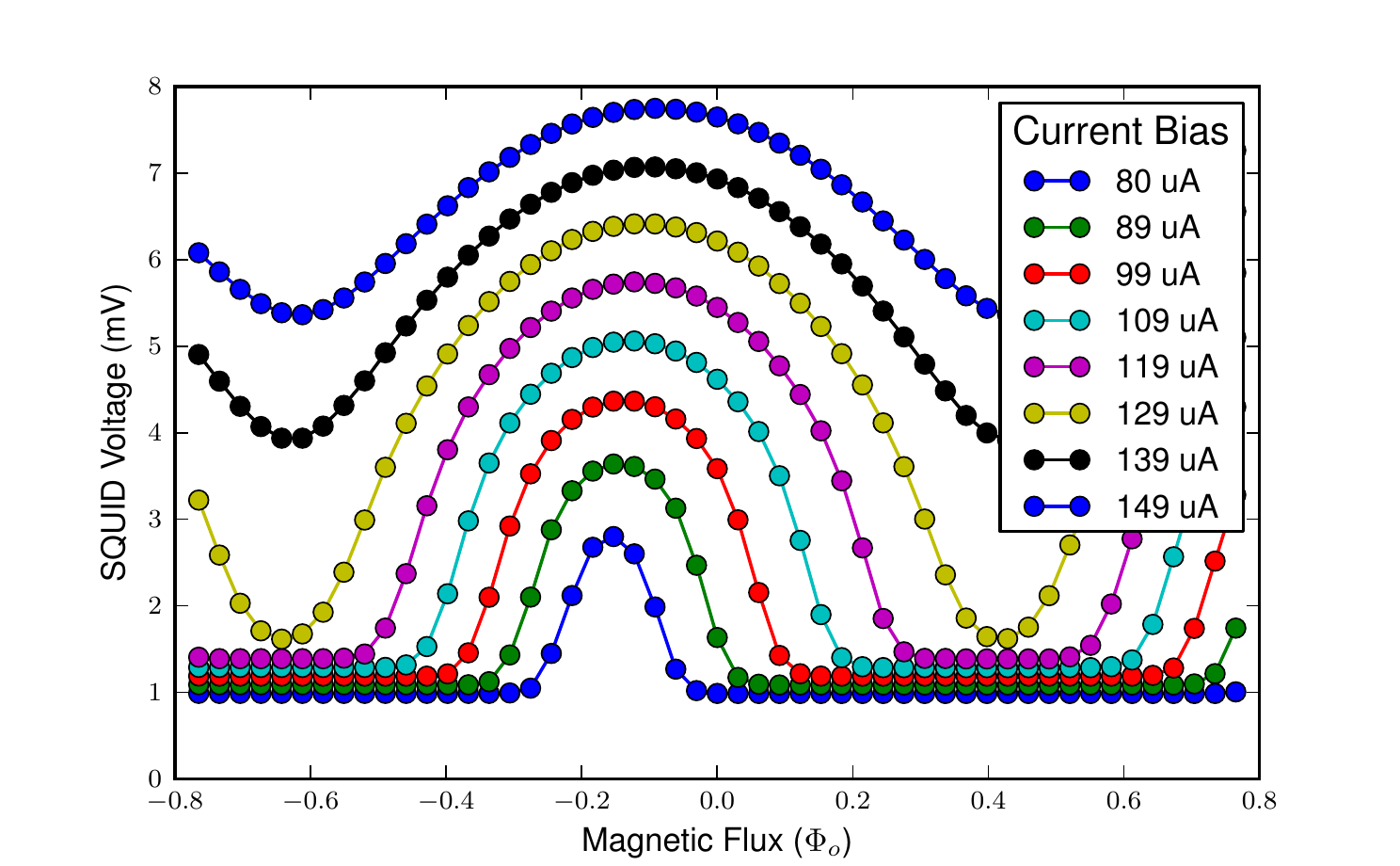}
    \caption{A set of SQUID V-$\Phi$ curves taken during the characterization of the EBEX SQUIDs. The magnetic flux is given in units of the magnetic flux quantum, $\Phi_o$. The current bias for each curve is provided in the legend. Note that the peak-to-peak voltage response increases with the first six current biases, but starts to decrease for current biases above 129$\mu$A.}
    \label{fig:squid_v_phis}
\end{center}
\end{figure}

Figure \ref{fig:squid_v_phis} shows the voltage response of the SQUID, $V_\mathrm{squid}$, as a function of magnetic 
flux. $\Phi$, and current bias, $I_\mathrm{bias}$. We refer to each curve in Figure~\ref{fig:squid_v_phis} 
as a ``V-$\Phi$'' curve.  Tuning the SQUID involves selecting the operating current bias such that the V-$\Phi$ has 
sufficient gain and dynamic range, and setting a fixed current current to the inductor coil (called the ``flux bias'') such that $\Phi$ corresponds to a 
section of the V-$\Phi$ with large negative slope. Once these operating parameters have been determined the SQUID 
is operated in shunt-feedback with a feedback resistance, $R_\mathrm{feedback}$ of 5~k\ohms. The procedure used 
to determine these parameters follows.

\begin{figure}[htbp]
\begin{center}
    \includegraphics[width=5in]{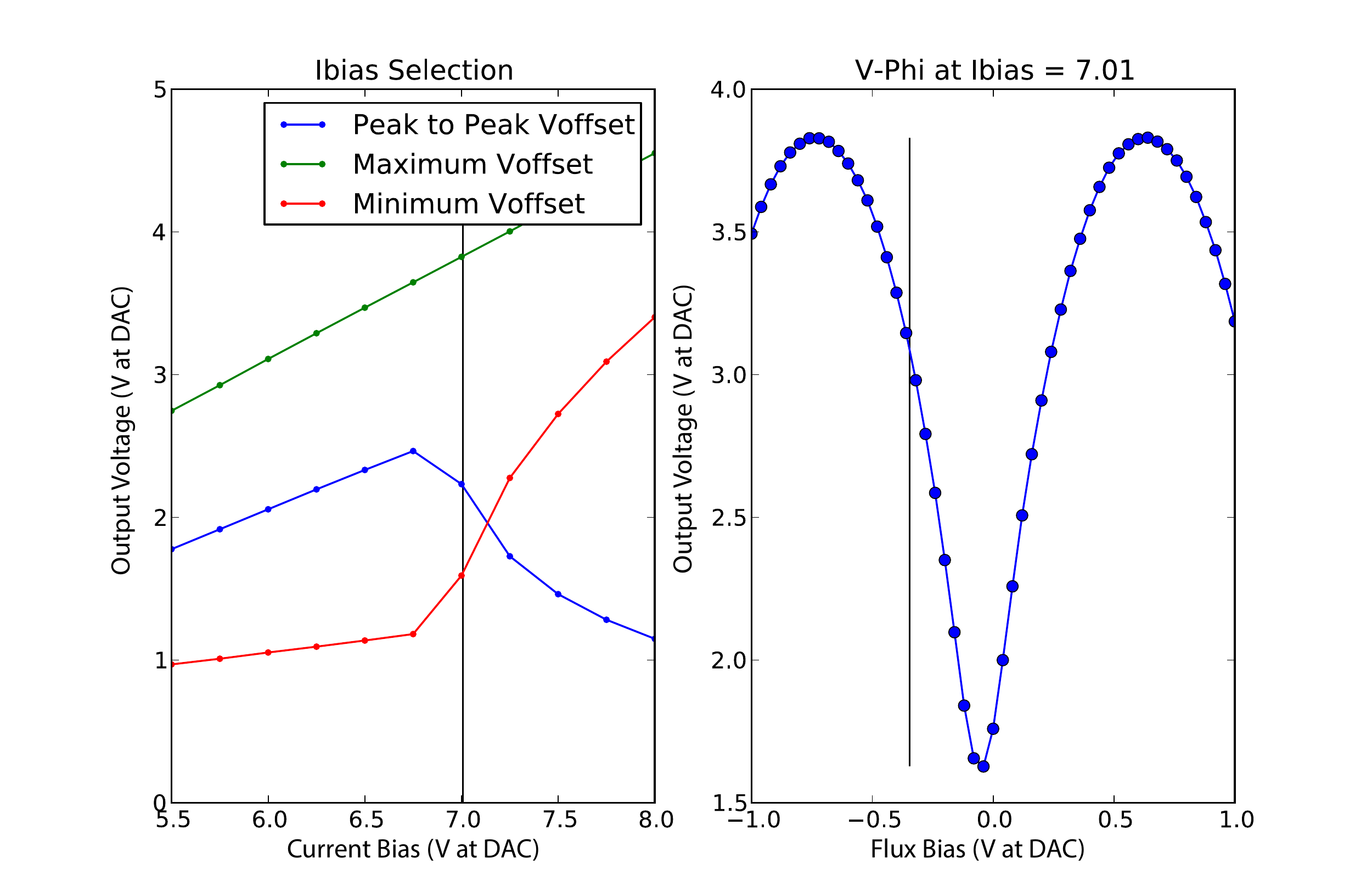}
    \caption{A successful SQUID tuning plot from the EBEX 2012/2013 science flight. These plots were generated in real time to assess the tuning state, and are in system units. The left plot shows the selection of the operating current bias. The green and red curves are the minimum and maximum voltage response of the SQUID at each current bias. The blue curve is the difference, and the vertical black line is selected current bias. The right plot is a V-$\Phi$ measured at the operating current bias. It is used to determine the operating flux bias. The selected flux bias is indicated by the vertical black bar.}
    \label{fig:example_squid_tune}
\end{center}
\end{figure}

The first step in the SQUID tuning process is the determination of the current bias. In practice, the tuning time is
reduced by taking as few measurements of $V_\mathrm{squid}$ as possible. Fortunately, the location of the peaks in each V-$\Phi$ are not
a function of current bias, so several measurements of $V_\mathrm{squid}$ per current bias are sufficient. The left panel of
Figure \ref{fig:example_squid_tune} shows data from this process. A reference V-$\Phi$ is measured (not shown) and used to
determine the flux biases corresponding to the peak and trough of SQUID voltage. For each current bias, the SQUID voltage
is measured only at several points near these flux biases. A parabolic fit is used to determine the maximum and minimum
 $V_\mathrm{sq}$ at this current bias. The results for each current bias are labelled ``Maximum Voffset'' and ``Minimum Voffset'' in Figure \ref{fig:example_squid_tune}. The peak-to-peak response is the difference between the maximum and minimum responses. The
  current bias selected is that with 90\% of the peak response. \revTwo{This selection reduces the gain of the SQUID slightly but has been empirically found to be more linear and stable.} The peak-to-peak response is labelled ``Peak to Peak Voffset'' on Figure
   \ref{fig:example_squid_tune}, the selected current bias is shown by the vertical black bar.

The second step is to determine the operating flux bias. A second V-$\Phi$ is measured at the current bias selected
during the first step, as shown in the right panel of figure \ref{fig:example_squid_tune}. The flux bias is selected to
be the average between the central flux bias, and the central output voltage. This point has been empirically determined
to be more stable than either of the central points alone. The right plot of Figure \ref{fig:example_squid_tune} shows
this process. The bias point is shown as a vertical black bar.

Table \ref{tab:ldb_squid_tunes} is a list of the seven tunings of the SQUID array involved with the 2012/2013 flight.
 The first tuning was performed on the ground prior to the flight. The full SQUID array consisted of 112 SQUIDs, of which 6 were known to have cryogenic wiring issues before flight. That left 106 SQUIDs which tuned successfully in tuning 1, prior to flight. During flight there were issues with the azimuthal pointing which resulted in the solar
panels not continuously pointing sun-ward. The reduction in power generation meant that only a section of the readout
was continuously powered. In addition, a secondary power issue developed near the end of flight that made it
impossible to tune half the SQUID array. This power issue was correlated with EBEX pointing within 40\degrees{} of the sun in azimuth
for nine hours, on January 6th, 2013. Therefore we attempted to tune only 52 SQUIDs on tunings 6 and 7; see 
Table~\ref{tab:ldb_squid_tunes}.

\begin{table}[htdp]
\begin{center}
\begin{tabular}{| c | c | c |}
\hline
Tuning & Start of  & Number of  \\
Number & Tuning (NZDT) & SQUIDs Tuned \\
\hline
1 & 2012-12-28 11:59:45 & 106 \\
2 & 2012-12-30 01:36:25 & 54 \\
3 & 2013-01-01 00:27:38 & 79 \\
4 & 2013-01-03 14:25:25 & 85 \\
5 & 2013-01-05 08:24:07 & 85 \\
6 & 2013-01-07 09:05:51 & 52 \\
7 & 2013-01-09 04:46:00 & 52 \\
\hline
\end{tabular}
\caption{The times and yields of each of the seven tunings of the 2012/2013 flight. Tuning 1 was performed on 
the ground before launch. In flight, fewer than the complete 112 SQUID array were attempted due to issues with
 the azimuthal pointing and power supplies, which resulted in powering off sections of the readout. \revTwo{The
  SQUID tuning procedure reliably tuned all the powered SQUIDs without known wiring issues during each tuning, 
  with the exception of one SQUID which showed intermittent tuning difficulties. It is not clear whether these
  issues are due to wiring or the tuning process.}
}
\label{tab:ldb_squid_tunes}
\end{center}
\end{table}

\begin{figure}[htbp] 
\begin{center}
\includegraphics[width=10cm]{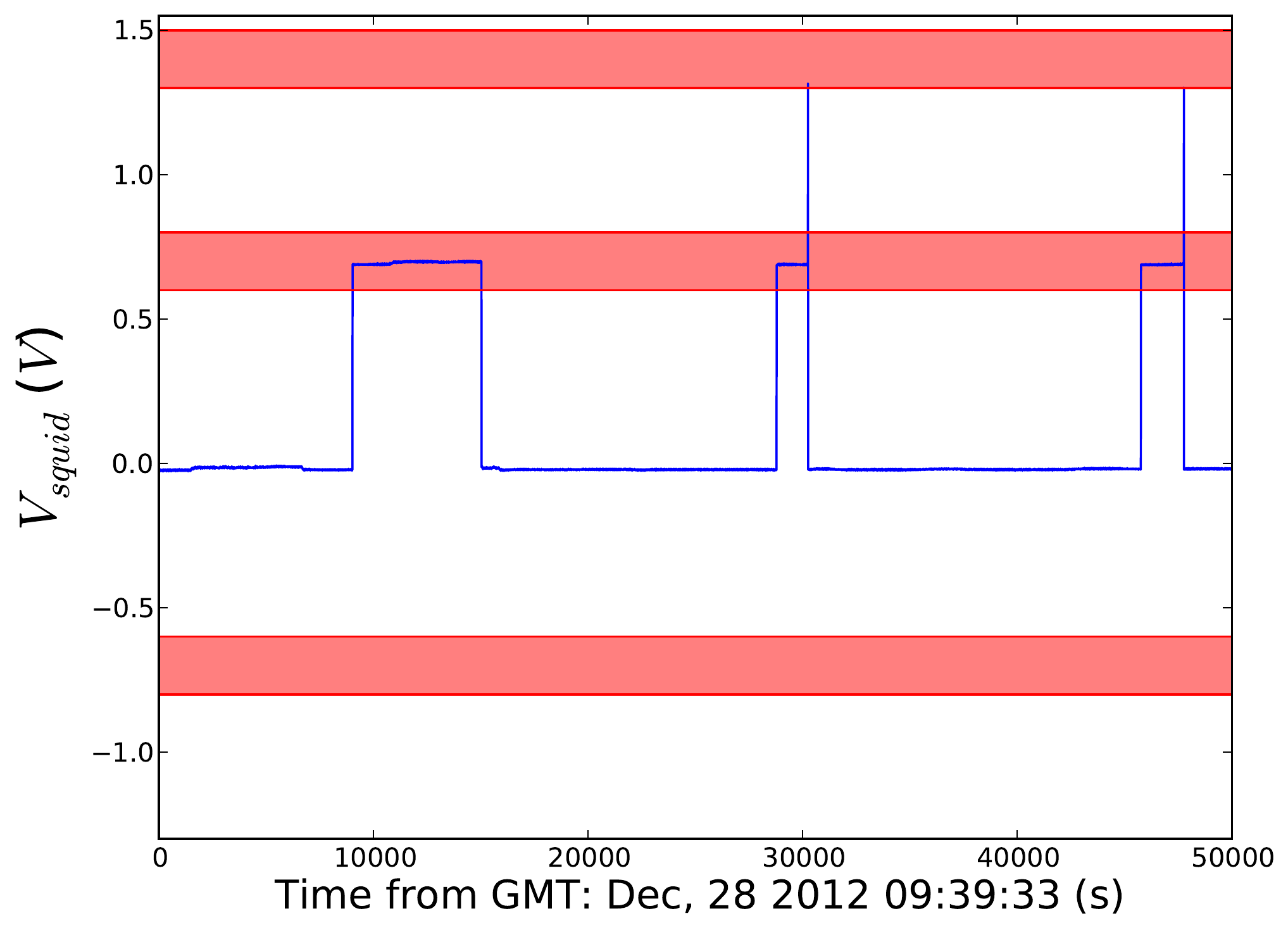}
\end{center}
\caption{
A timestream of SQUID during the EBEX LDB flight selected for its high number of flux jump events. Flux jumps are easily
detected when the DC out of the SQUID becomes an integral multiple of $\pm$ 0.7 $V$. This timestream shows three flux
jump events. Each is corrected by a deflux jump tuning algorithm that involves adjusting the flux bias until the SQUID
returns to its original operating point. This procedure often temporarily introduces an increase in $V_{squid}$, as
shown in the second and third event. These spikes are likely caused by the flux bias adjustment causing the SQUID
to jump another flux quantum. The deflux jump algorithm is configured to retry in this case, and the majority of cases
successfully returns the SQUID to its original operating point.
}
\label{fig:example_squid_ts_with_flux_jumps}
\end{figure}

Another measure of SQUID stability is the measurement SQUID DC voltage during observations. SQUIDs can jump an integer
number of flux quanta, which causes the output voltage to step by a fixed voltage, roughly 0.7~V at the ADC of our
system. While flux jumped, the SQUID still operates but has a significantly reduced 
dynamic range. An example SQUID DC timestream with a much higher than average number of flux jumps is shown in
Figure~\ref{fig:example_squid_ts_with_flux_jumps}. Analysis of the SQUID DC voltage over the whole flight shows that
they were flux jumped for less than 2\% of the observation time.

\subsubsection{Bolometer Tuning}
\label{sec:bolometer tuning}

Tuning the bolometer array was a two step process, consisting of first ``overbiasing'' the detector, supplying sufficient
 electrical power to warm the TES above its critical temperature, and then ``biasing'' the detector, by dropping this
  electrical power such that the TES enters its superconducting transition. The overbias must be applied when the
   bolometer is in a non-superconducting state, so it was often done near the end of the fridge cycle procedure, while the  temperature of the bolometers were roughly 0.8-1~K.

\begin{figure}[htbp]
	\begin{centering}
    \includegraphics[width=6in]{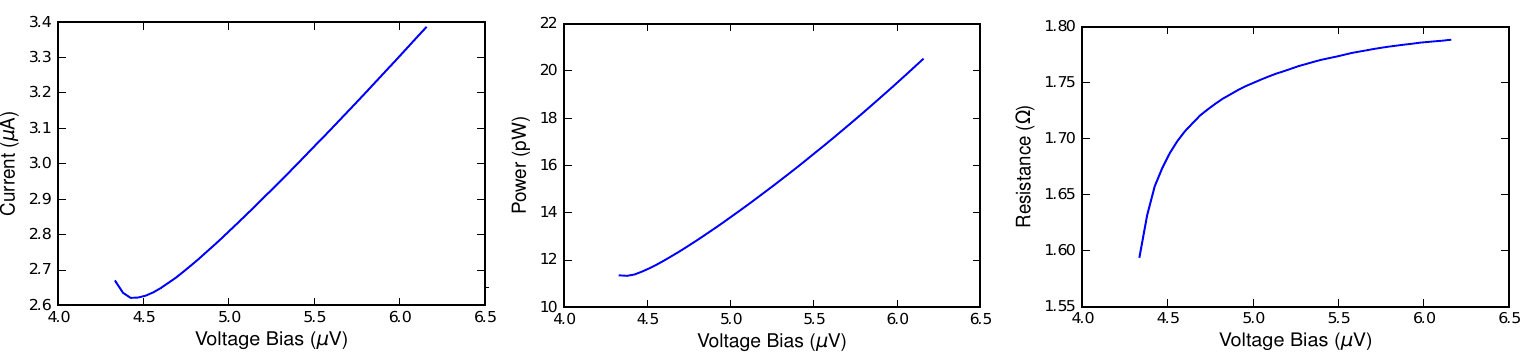}
    \caption{
    Example current-voltage (IV), power-voltage (PV) and resistance-voltage (RV) curves for a 410~GHz bolometer tuned
during flight. The turnaround in the IV and flatness of the PV at low voltage are indication that this channel is
exhibiting negative electro-thermal feedback.}
    \label{fig:bolo_iv}
    \end{centering}
\end{figure}

\revOne{Biasing a bolometer into the transition involves lowering the bias voltage until the TES reaches a fixed fraction of its
overbiased resistance, typically 85\% for EBEX. This fraction has been found to provide stable operation and adequate loop gain. Figure~\ref{fig:bolo_iv} shows example current-voltage (IV), power-voltage (PV),
and resistance-voltage (RV) curves for a single channel during the first tuning of flight. The turnaround in the IV at
low voltage bias indicates that the TES has entered the transition and is exhibiting negative electro-thermal feedback.
The most common bolometer tuning issue encountered during flight was bolometer saturation. In this case, the total
optical power deposited on the detector is sufficient to keep the TES above its critical temperature even in the absence
of electrical power. This is evident in the PV curve of the detector, as the electrical power supplied drops to zero.
}

The bolometer array was tuned 21 times during the flight, many more than the 6 times required by fridge cycles. 
This was mainly due to measurements of the optical loading from the atmosphere, which involved retuning the
bolometers at several telescope elevation angles. The change in optical load can be determined from the change in power in the PV curve measured as part of the tuning process. These measurements were performed three times during the flight, each involved four to five bolometer tuning procedures.

\revOne{The only time during the flight that tuning of the full bolometer array was attempted was the first tuning, done shortly after ascent. The results of this tuning were that 955 of the 1190 wired channels had turnarounds in their IV curves indicating negative electro-thermal feedback. Of the remaining 235, 133 were saturated, 58 had unexpected IV curves, 22 had a temporary tuning issue and 22 were resistor channels.  That some channels would saturate in flight was expected, as precise optical loading
measurements were not available prior to flight, so we intentionally selected detector wafers with a range of saturation powers. The most likely cause of the unexpected IV curves is cross talk between channels during the tuning procedure. This had the largest effect on the first tuning, where the overbias amplitudes were conservatively set several times higher than required. The measurements from this tuning were used to individually adjust the overbias amplitudes for each channel to be approximately 10\% above the transition region, minimizing the effect of cross talk on the tuning procedure. The temporary tuning issue involved one mezzanine board not powering correctly during this first tuning, it was powered for subsequent tunings. }

\revOne{The total observation time for each frequency band were 26844, 19808, 8064 bolometer hours for 150, 250 and
 410~GHz, respectively. These were calculated by constructing a flight timeline for each detector from the results of each of the
 21 tunings and the flight control computer logs. This corresponds to approximately 25\% of the observation time possible if the entire readout was continuously powered.}

\subsection{Noise}
\revOne{
This section discusses the noise of the readout system and bolometers during one early segment of the
flight, on December 31st, 2012. This segment has been the primary focus of the EBEX analysis so far due to its high number of tuned detectors
and length.
}

\begin{figure}[htbp]
    \begin{center}
    \includegraphics[width=5in]{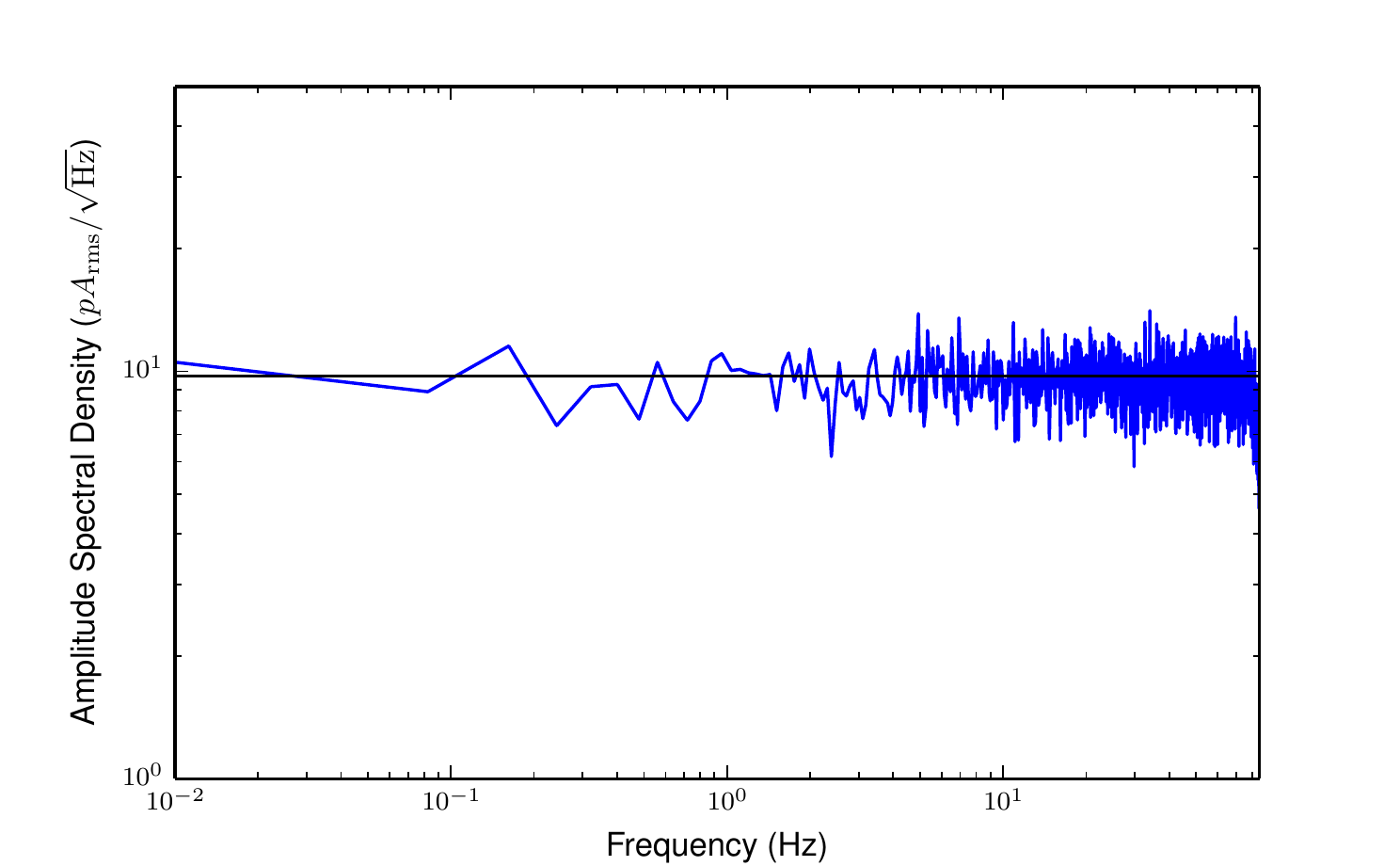}
    \caption{The current noise for dark SQUID B206 demodulated at 915057~$\hertz$. The horizontal line is the predicted dark SQUID noise. The noise measured in the EBEX optical band is 9.23~${pA_\mathrm{rms}\over\sqrt{Hz}}$.
    The Nyquist frequency is 95~Hz. }
    \label{fig:dark_squid_noise}
    \end{center}
\end{figure}

\begin{figure}[htbp]
    \begin{center}
    \includegraphics[width=5in]{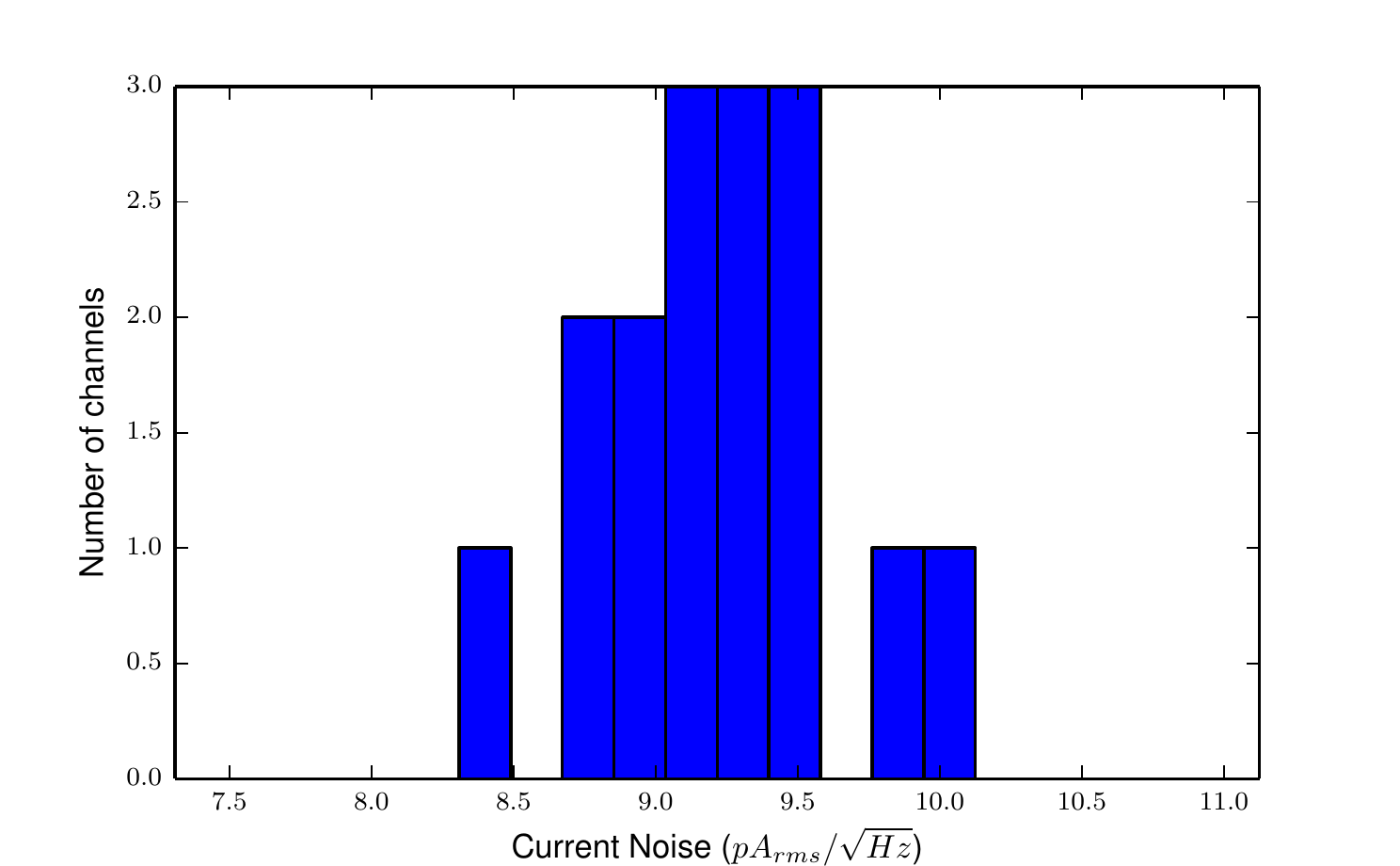}
    \caption{The current noise of the 16 frequency windows on dark SQUID B206. The dark SQUID noise prediction, shown on figure \ref{fig:dark_squid_noise}, is 9.75~${pA_\mathrm{rms} \over \sqrt{Hz}}$, consistent with these measurements given the expected variation in readout components.}
    \label{fig:dark_squid_histogram}
    \end{center}
\end{figure}

EBEX included two ``dark'' SQUID channels, which were identical to the other SQUID channels except that the connections to the bolometer circuit were left open. These channels act as a test of the SQUID and readout noise. Each dark 
SQUID was demodulated at 16 frequencies between 
400 and 1200~kHz. The frequencies and demodulation
bandwidths were similar to those used for bolometer channels.
Figure \ref{fig:dark_squid_noise} shows the amplitude spectral density for one of the dark SQUIDs in a frequency window at 
915~kHz. Figure \ref{fig:dark_squid_histogram} is a histogram of the ratio of measured to predicted noise for all 16
 frequency windows on this dark SQUID. It corresponds to the average noise in the amplitude spectral 
 density from 2.90 to 4.80~Hz.
This is the bandwidth of polarized signals for light detectors. The noise for both dark SQUIDs is consistent with predictions.

We use well established noise formulas to calculate noise predictions for bolometers open to light~\cite{Mather1982}.
The predicted contributions from each of the main noise sources for example channels in each frequency band are given in 
Table~\ref{tab:noise_three_channel}. The parameters used in the calculation are given in 
Table~\ref{tab:noise_params_channel}. The third column, ``Source,'' indicates which parameters were measured during preflight testing (``Preflight''), during flight (``Flight''), and those which were assumed (``Assumed''). The optical loading $P_\mathrm{opt}$ is
 determined by differencing the electrical power required to bias the detector during flight with that measured during preflight tests in which steps were taken to reduce the optical loading to negligible levels. The readout noise depends on the gain setting of the multiplexed module of the channel, as well as the voltage bias supplied. As such, it varies slightly from channel to channel.

The scaling of several noise sources relies on the responsivity of the bolometer, $S_{i} = dI/dP$. For a bolometer
exhibiting negative electro-thermal feedback
\begin{equation}
S_{i} = {-\sqrt{2} \over V^\mathrm{RMS}_\mathrm{bias}} \left({\loopg \over \loopg + 1} \right) { 1 \over 1 + i \omega \tau },
\label{eq:responsivity}
\end{equation}
where $V^\mathrm{RMS}_\mathrm{bias}$ is the voltage bias on the bolometer, $\loopg$ is the loop gain, $\omega$ is the
angular frequency of the signal, and $\tau$ is the electro-thermal time constant of the bolometer.
For Table \ref{tab:noise_params_channel} the voltage bias is measured as part of the IV curve and the loop gain is fixed
to 10, an estimate based on lab tests. The noise sections presented are within the bolometer signal band, where $\omega \tau \ll 1$.

\revOne{
Note that the predicted current noise at 150~GHz exceeds that at 410~GHz. This is partially due to the higher than
expected optical load in this band. The expected load at 150~GHz was 2.2~pW, whereas the measured load was 5.7 pW for this channel.  The average optical loading for all 150~GHz channels was measured to be 3.6~pW during flight, just over 60\% higher than expected. For the 250~GHz and 410~GHz the average measured optical loads are 5.1~pW and 4.8~pW, respectively. These are reasonable given their 7 and 12~pW respective predicted optical loads. More detailed analysis of the optical loads is ongoing.} 


\begin{table}[htdp]
\begin{center}
\begin{tabular}{| c | c | c | c | c |}
\hline
Noise Source & Equation & 150~GHz & 250~GHz & 410~GHz \\
& & (${pA_\mathrm{rms}\over \sqrt{\hertz}}$) & (${pA_\mathrm{rms} \over \sqrt{\hertz}}$) & (${pA_\mathrm{rms} \over \sqrt{\hertz}}$) \\
\hline
Readout & N/A & 13 & 12 & 14 \\
Johnson &  $ \sqrt{2} \cdot \sqrt{4 k T_c \over R}$ & 8 & 8 & 8 \\
Phonon & $ S_i \cdot \sqrt{4 \gamma k T_c^2 G}$& 15 & 15 & 12 \\
Photon & $ S_i \cdot \sqrt{2h \nu P_\mathrm{opt}}$ & 24 & 29 & 18 \\
Photon Bunching & $ S_i \cdot \sqrt{P_\mathrm{opt}^2 \over \Delta \nu}$ & 26 & 23 & 7 \\
\hline
Total & & 42 & 43 & 28 \\
\hline
\end{tabular}
\caption{
The predicted noise for a typical bolometer channel from each frequency band. 
The three channels shown are 150-47-05-03,
250-23-04-08, and 410-28-10-02. 
The properties for each channel are given in Table \ref{tab:noise_params_channel}. The
$\sqrt{2}$ in the Johnson noise is due to the contribution from both carrier side bands. 
Details can be found in Appendix A of Dobbs, 2012~\cite{Dobbs2012_readout}. The noise predictions have a 35\% uncertainty due primarily to uncertainty in the measured optical load, applied voltage bias and spectral band width. 
}
\label{tab:noise_three_channel}
\end{center}
\end{table}
\begin{table}[htdp]
\begin{center}
\begin{tabular}{| c | c | c | c | c | c | }
\hline
& & & \multicolumn{3}{| c |}{Value} \\
Property & Symbol & Source & 150~GHz & 250~GHz & 410~GHz \\
\hline
Critical Temperature (mK) & $T_c$ & Preflight & 560 & 480 & 510 \\
Bolometer Resistance ($\Omega$) & $R$ & Flight & 1.0 & 0.94 & 0.78 \\
Thermal Conductivity (pW/K) & $G_\mathrm{dyn}$ & Preflight & 54 & 93 & 140 \\
Band Center (GHz) & $\nu$ & Preflight & 150 & 250 & 400 \\
Band Width (GHz) & $\Delta \nu$ & Preflight & 23 & 33 & 51 \\
Optical Power (pW) & $P_{opt}$ & Flight/Preflight & 5.7 & 7.1 & 4.4 \\
Loop gain & $\loopg$ & Assumed & 10 & 10 & 10 \\
Responsivity (kA/W) & $S_i$ & Flight & 700 & 602 & 370 \\
\hline
\end{tabular}
\caption{The operating parameters that are used to predict the noise 
for example bolometers 150-47-05-03. 250-23-04-08, and 410-28-10-02. 
The third column lists the source of each parameter, either ``Preflight'' for preflight calibration tests,
 ``Flight'' for parameters 
measured as part of the tuning process, or ``Assumed'' for assumed values. The optical power is listed as ``Flight/Preflight'' as it is the difference between the preflight measurement of saturation power in a dark cryostat and that measured during flight.  There is a 5-10\% uncertainty on most
 of these parameters, with the exception of the responsivity and optical power, which are known to 20\%, the band
  width, which is currently known to roughly 30\%, and the loop gain, which is unmeasured in flight. To calculate the noise uncertainty a loop gain range from 3 to 20 is used, as an estimate from typical TES operation \cite{Dobbs2012_readout}. The
   responsivity, optical power, and band width contribute significantly to the uncertainty in the noise prediction,
    the loop gain has a less significant effect, due to the ${ \loopg \over \loopg + 1}$ dependence of Equation \ref{eq:responsivity}.}

\label{tab:noise_params_channel}
\end{center}
\end{table}

\begin{figure}[htbp]
    \begin{center}
    \includegraphics[width=6in]{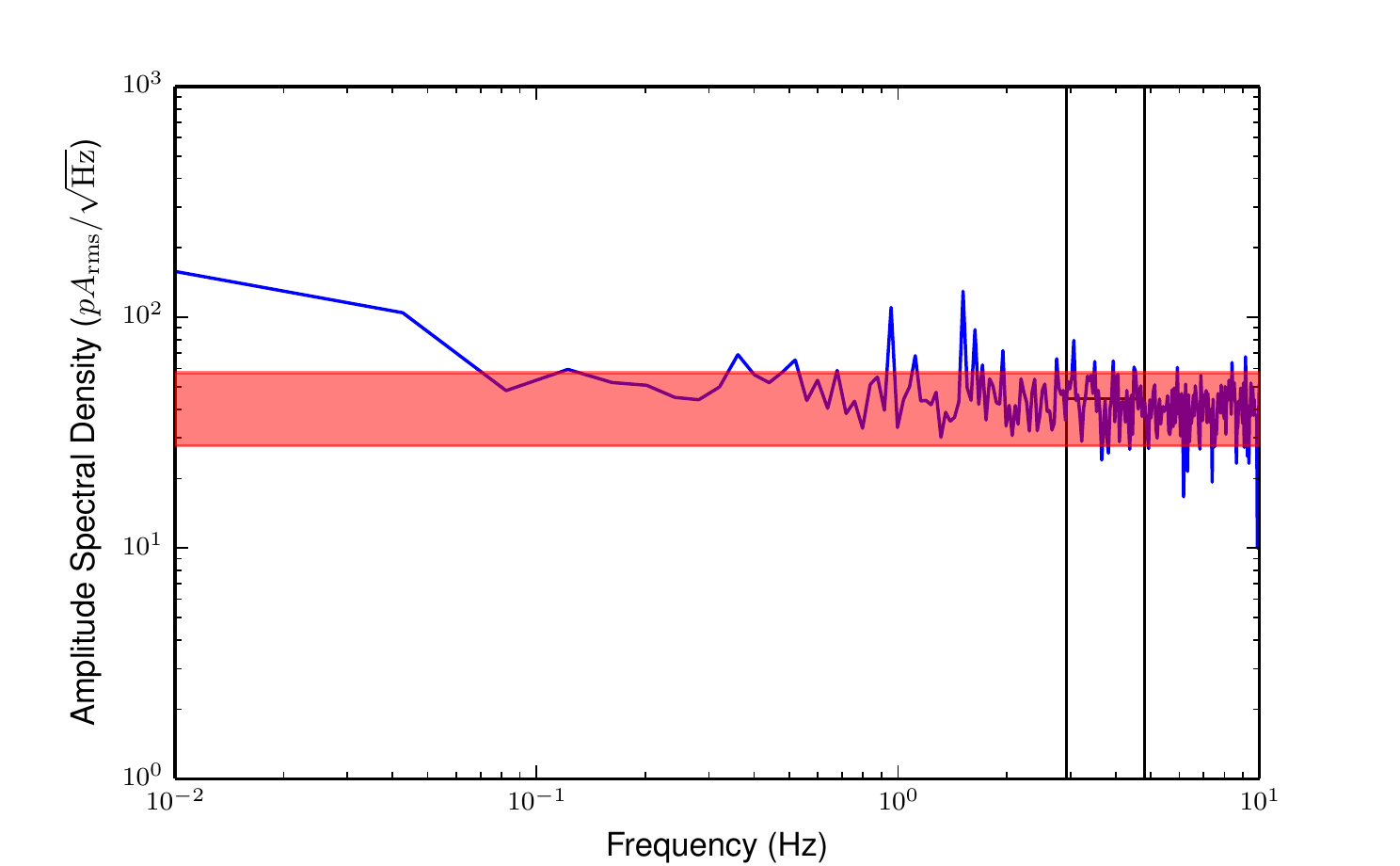}
    \caption{The current noise for the 250~GHz detector 250-23-04-08. The horizontal band is the predicted 
    noise given in Table~\ref{tab:noise_three_channel} with a 35\% error band due primarily to uncertainty in the spectral band, voltage bias, and the optical loading.
    The vertical lines demarcate the optical band for polarized signals, from 2.9 to 4.8~Hz. The average noise is shown as a short horizontal line in this section. It agrees with the prediction, given the uncertainty.}
    \label{fig:light_channel_noise}
    \end{center}
\end{figure}

\begin{figure}
  \centering
  \includegraphics[width=10cm]{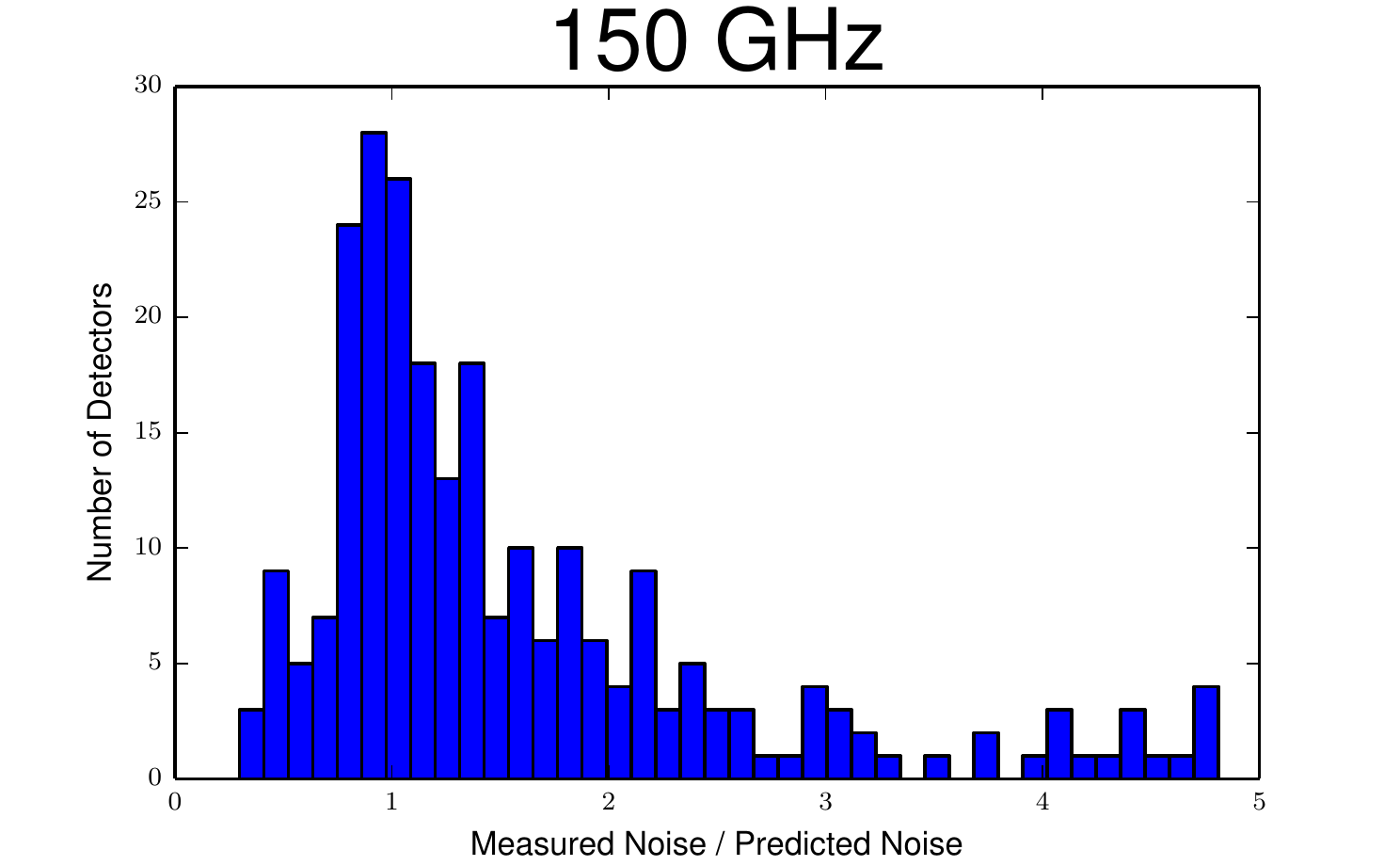}
  \caption{The measured to predicted noise ratio for 270 150~GHz channels tuned on December 31st, 2012. 
  Twenty-three channels showed noise greater than 5 times the prediction and are not shown here.}
  \label{fig:noise_hist:150}
\end{figure}

\begin{figure}
  \centering
  \includegraphics[width=10cm]{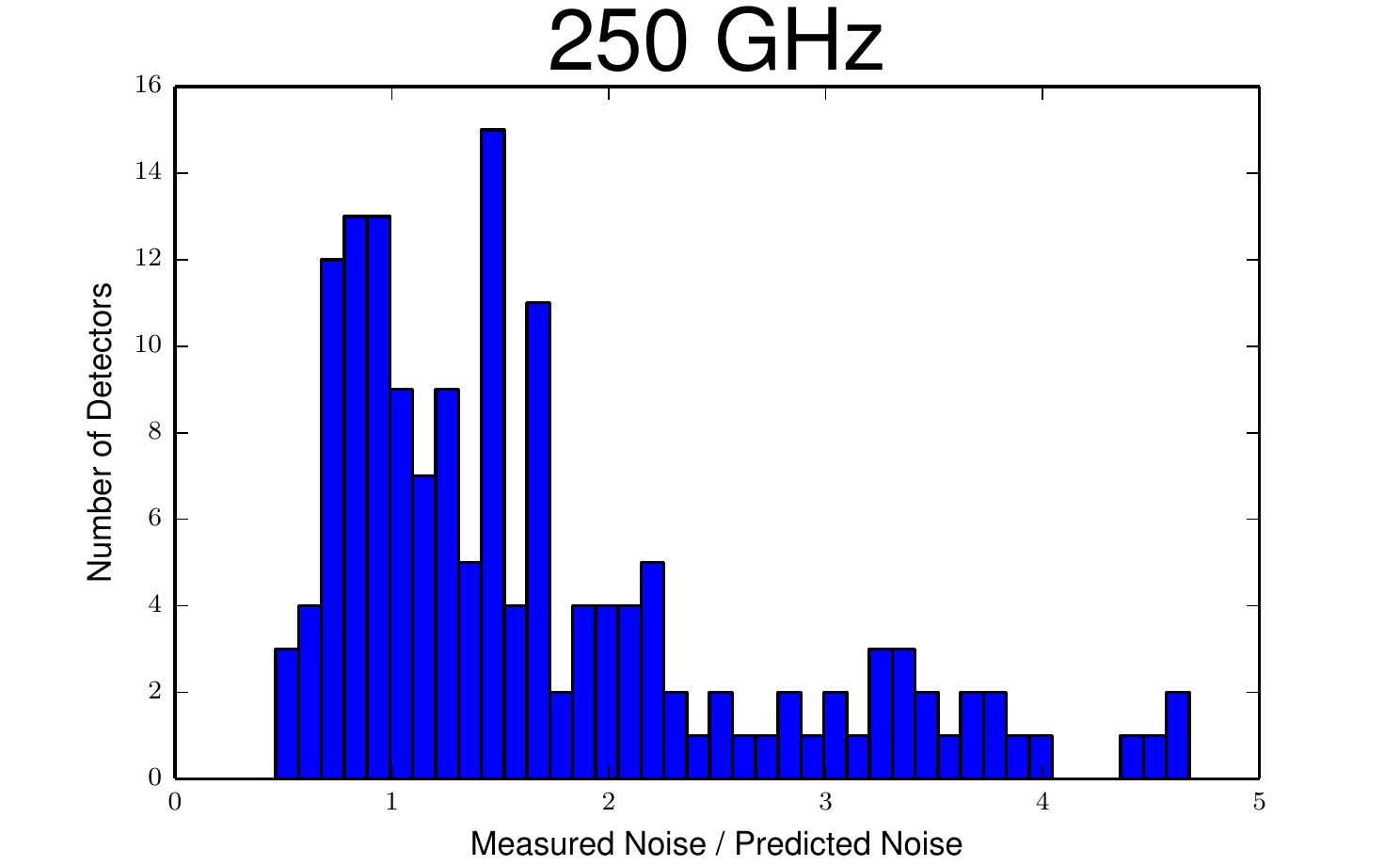}
  \caption{The measured to predicted noise ratio for 186 250~GHz channels tuned on December 31st, 2012. 
  Thirty channels showed noise greater than 5 times the prediction and are not shown here.}
  \label{fig:noise_hist:250}
\end{figure}

\begin{figure}
  \centering
  \includegraphics[width=10cm]{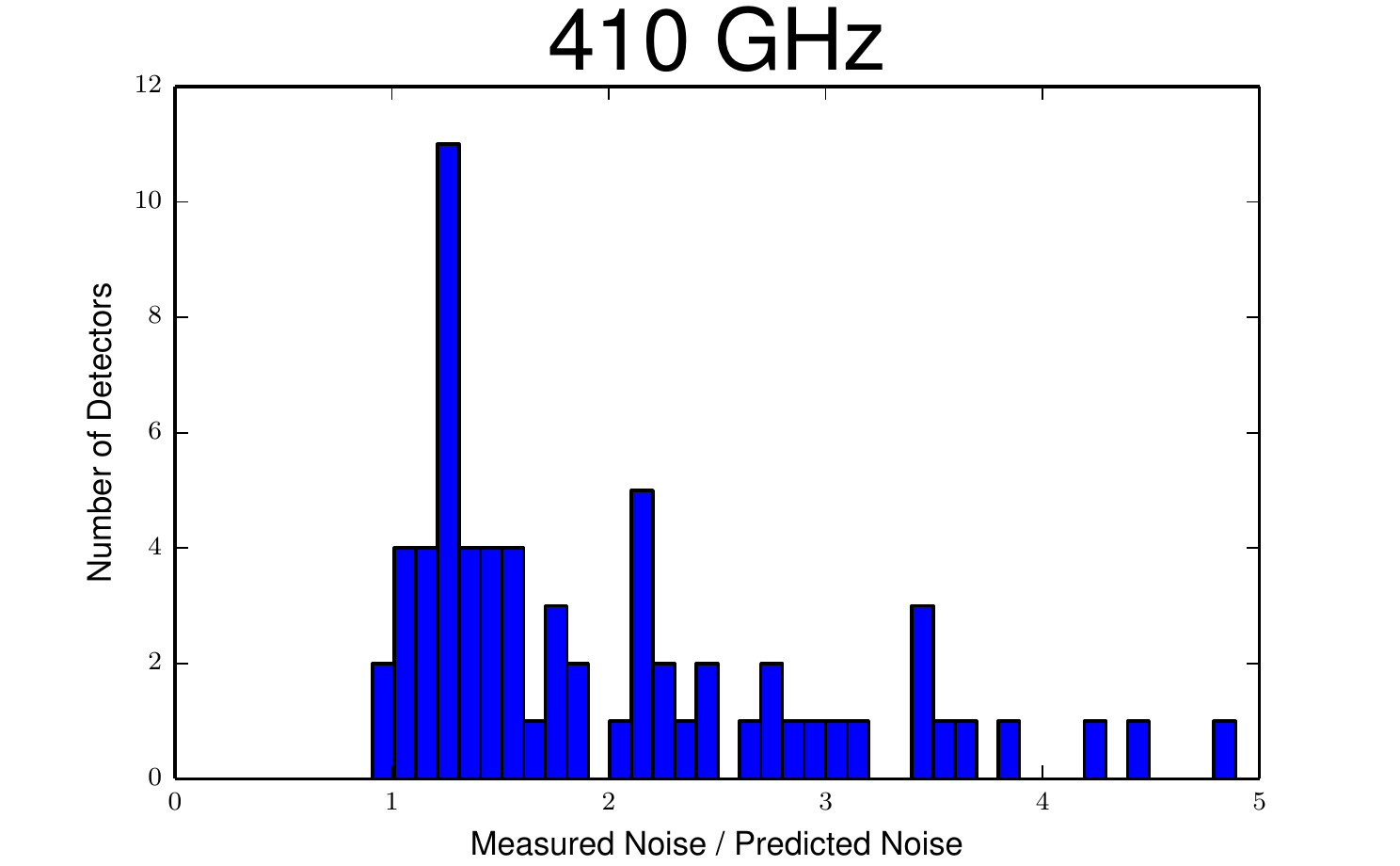}
  \caption{The measured to predicted noise ratio for 70 410~GHz channels tuned on December 31st, 2012. 
  Four channels showed noise greater than 5 times the prediction and are not shown here.}
  \label{fig:noise_hist:410}
\end{figure}

\revTwo{The same calculations shown for three example channels in Table~\ref{tab:noise_three_channel} and 
Table~\ref{tab:noise_params_channel} were repeated for every channel tuned on December 31st,
2012. In addition, noise for each was measured by calculating the power spectral density and averaging a
window from 2.9 to 4.8~Hz, as shown in figure~\ref{fig:light_channel_noise}. 
This frequency window corresponds to the optical band for polarized signals, though the sky signals present in these detectors are below the noise level.
Figures~\ref{fig:noise_hist:150},~\ref{fig:noise_hist:250}, and~\ref{fig:noise_hist:410} show the noise ratios for the
tuned channels in the 150, 250, and 410~GHz bands, respectively. 
The majority of bolometers operated at current 
noise level at or below 1.4, 1.6 and 1.8 for the 150, 250 and 410~GHz bands, respectively. Further
evaluation of the bolometer performance during the flight is ongoing. In particular, noise equivalent temperatures are not presented here as analysis  
of the telescope beam, calibration, and spectral bands is in progress.}

\section{Conclusion}

\revTwo{EBEX successfully operated kilo-pixel array of TES bolometers in a space-like environment during its eleven day flight over the Antarctic in 2012, 2013. Several challenges involved in operating the digital frequency domain readout electronics and SQUID amplifiers on the balloon platform were recognized prior to flight and addressed. Bolometer and SQUID tuning during the flight, performed remotely with little communication with the payload, have been shown to be reliable and provide high yield, despite additional challenges presented by azimuthal pointing issues unrelated to the bolometers and readout.}

\revTwo{The current noise of both the dark SQUIDs and light bolometer channels have been measured during an early section of flight. Noise predictions taking into account preflight and flight instrument properties agree with these measurements, indicating that the bolometers and readout are behaving as expected in the flight environment.}

\revTwo{EBEX data analysis is ongoing, including measuring the noise for larger sections of flight, and refining measurements and analysis of the beam, spectral bands, and calibration to permit the calculation of the bolometer noise equivalent temperatures.}

\section{Acknowledgements}
EBEX is a NASA supported mission through grant numbers NNX08AG40G and NNX07AP36H. We thank
Columbia Scientific Balloon Facility for their enthusiastic support of EBEX. We also acknowledge support from
NSF, CNRS, Minnesota Super Computing Institute, Minnesota and Rhode Island Space Grant Consortia, the
Science and Technology Facilities Council in the UK, Sigma Xi, and funding from collaborating institutions.
This research used resources of the National Energy Research Scientific Computing Center, which is supported by
the office of Science of the U.S. Department of Energy under contract No. DE-AC02-05CH11231. The McGill
authors acknowledge funding from the Canadian Space Agency, Natural Sciences and Engineering Research
Council, Canadian Institute for Advanced Research, Canadian Foundation for Innovation and Canada Research
Chairs program.

\bibliography{2014_spie_readout}   

\begin{thebibliography}{10}

\bibitem{Aubin2010}
F.~Aubin, A.~M. Aboobaker, P.~Ade, C.~Baccigalupi, C.~Bao, J.~Borrill,
  C.~Cantalupo, D.~Chapman, J.~Didier, M.~Dobbs, W.~Grainger, S.~Hanany,
  J.~Hubmayr, P.~Hyland, S.~Hillbrand, A.~Jaffe, B.~Johnson, T.~Jones,
  T.~Kisner, J.~Klein, A.~Korotkov, S.~Leach, A.~Lee, M.~Limon, K.~MacDermid,
  T.~Matsumura, X.~Meng, A.~Miller, M.~Milligan, D.~Polsgrove, N.~Ponthieu,
  K.~Raach, B.~Reichborn-Kjennerud, I.~Sagiv, G.~Smecher, H.~Tran, G.~S.
  Tucker, Y.~Vinokurov, A.~Yadav, M.~Zaldarriaga, and K.~Zilic, ``First
  implementation of tes bolometer arrays with squid-based multiplexed readout
  on a balloon-borne platform,'' {\em Proc. SPIE}~{\bf 7741},
  pp.~77411T--77411T--10, 2010.

\bibitem{Polarbear}
Z.~D. Kermish, P.~Ade, A.~Anthony, K.~Arnold, D.~Barron, D.~Boettger,
  J.~Borrill, S.~Chapman, Y.~Chinone, M.~A. Dobbs, J.~Errard, G.~Fabbian,
  D.~Flanigan, G.~Fuller, A.~Ghribi, W.~Grainger, N.~Halverson, M.~Hasegawa,
  K.~Hattori, M.~Hazumi, W.~L. Holzapfel, J.~Howard, P.~Hyland, A.~Jaffe,
  B.~Keating, T.~Kisner, A.~T. Lee, M.~Le~Jeune, E.~Linder, M.~Lungu,
  F.~Matsuda, T.~Matsumura, X.~Meng, N.~J. Miller, H.~Morii, S.~Moyerman, M.~J.
  Myers, H.~Nishino, H.~Paar, E.~Quealy, C.~L. Reichardt, P.~L. Richards,
  C.~Ross, A.~Shimizu, M.~Shimon, C.~Shimmin, M.~Sholl, P.~Siritanasak,
  H.~Spieler, N.~Stebor, B.~Steinbach, R.~Stompor, A.~Suzuki, T.~Tomaru,
  C.~Tucker, and O.~Zahn, ``The polarbear experiment,'' {\em Proc. SPIE}~{\bf
  8452}, pp.~84521C--84521C--15, 2012.

\bibitem{Sptpol}
J.~E. {Austermann}, K.~A. {Aird}, J.~A. {Beall}, D.~{Becker}, A.~{Bender},
  B.~A. {Benson}, L.~E. {Bleem}, J.~{Britton}, J.~E. {Carlstrom}, C.~L.
  {Chang}, H.~C. {Chiang}, H.-M. {Cho}, T.~M. {Crawford}, A.~T. {Crites},
  A.~{Datesman}, T.~{de Haan}, M.~A. {Dobbs}, E.~M. {George}, N.~W.
  {Halverson}, N.~{Harrington}, J.~W. {Henning}, G.~C. {Hilton}, G.~P.
  {Holder}, W.~L. {Holzapfel}, S.~{Hoover}, N.~{Huang}, J.~{Hubmayr}, K.~D.
  {Irwin}, R.~{Keisler}, J.~{Kennedy}, L.~{Knox}, A.~T. {Lee}, E.~{Leitch},
  D.~{Li}, M.~{Lueker}, D.~P. {Marrone}, J.~J. {McMahon}, J.~{Mehl}, S.~S.
  {Meyer}, T.~E. {Montroy}, T.~{Natoli}, J.~P. {Nibarger}, M.~D. {Niemack},
  V.~{Novosad}, S.~{Padin}, C.~{Pryke}, C.~L. {Reichardt}, J.~E. {Ruhl}, B.~R.
  {Saliwanchik}, J.~T. {Sayre}, K.~K. {Schaffer}, E.~{Shirokoff}, A.~A.
  {Stark}, K.~{Story}, K.~{Vanderlinde}, J.~D. {Vieira}, G.~{Wang},
  R.~{Williamson}, V.~{Yefremenko}, K.~W. {Yoon}, and O.~{Zahn}, ``{SPTpol: an
  instrument for CMB polarization measurements with the South Pole
  Telescope},'' in {\em Society of Photo-Optical Instrumentation Engineers
  (SPIE) Conference Series},  {\em Society of Photo-Optical Instrumentation
  Engineers (SPIE) Conference Series} {\bf 8452}, Sept. 2012.

\bibitem{Dobbs2012_readout}
M.~A. {Dobbs}, M.~{Lueker}, K.~A. {Aird}, A.~N. {Bender}, B.~A. {Benson}, L.~E.
  {Bleem}, J.~E. {Carlstrom}, C.~L. {Chang}, H.-M. {Cho}, J.~{Clarke}, T.~M.
  {Crawford}, A.~T. {Crites}, D.~I. {Flanigan}, T.~{de Haan}, E.~M. {George},
  N.~W. {Halverson}, W.~L. {Holzapfel}, J.~D. {Hrubes}, B.~R. {Johnson},
  J.~{Joseph}, R.~{Keisler}, J.~{Kennedy}, Z.~{Kermish}, T.~M. {Lanting}, A.~T.
  {Lee}, E.~M. {Leitch}, D.~{Luong-Van}, J.~J. {McMahon}, J.~{Mehl}, S.~S.
  {Meyer}, T.~E. {Montroy}, S.~{Padin}, T.~{Plagge}, C.~{Pryke}, P.~L.
  {Richards}, J.~E. {Ruhl}, K.~K. {Schaffer}, D.~{Schwan}, E.~{Shirokoff},
  H.~G. {Spieler}, Z.~{Staniszewski}, A.~A. {Stark}, K.~{Vanderlinde}, J.~D.
  {Vieira}, C.~{Vu}, B.~{Westbrook}, and R.~{Williamson}, ``{Frequency
  multiplexed superconducting quantum interference device readout of large
  bolometer arrays for cosmic microwave background measurements},'' {\em Review
  of Scientific Instruments}~{\bf 83}, p.~073113, July 2012.

\bibitem{Britt2010}
B.~{Reichborn-Kjennerud}, A.~M. {Aboobaker}, P.~{Ade}, F.~{Aubin},
  C.~{Baccigalupi}, C.~{Bao}, J.~{Borrill}, C.~{Cantalupo}, D.~{Chapman},
  J.~{Didier}, M.~{Dobbs}, J.~{Grain}, W.~{Grainger}, S.~{Hanany},
  S.~{Hillbrand}, J.~{Hubmayr}, A.~{Jaffe}, B.~{Johnson}, T.~{Jones},
  T.~{Kisner}, J.~{Klein}, A.~{Korotkov}, S.~{Leach}, A.~{Lee}, L.~{Levinson},
  M.~{Limon}, K.~{MacDermid}, T.~{Matsumura}, X.~{Meng}, A.~{Miller},
  M.~{Milligan}, E.~{Pascale}, D.~{Polsgrove}, N.~{Ponthieu}, K.~{Raach},
  I.~{Sagiv}, G.~{Smecher}, F.~{Stivoli}, R.~{Stompor}, H.~{Tran},
  M.~{Tristram}, G.~S. {Tucker}, Y.~{Vinokurov}, A.~{Yadav}, M.~{Zaldarriaga},
  and K.~{Zilic}, ``{EBEX: a balloon-borne CMB polarization experiment},'' in
  {\em Society of Photo-Optical Instrumentation Engineers (SPIE) Conference
  Series},  {\em Society of Photo-Optical Instrumentation Engineers (SPIE)
  Conference Series} {\bf 7741}, July 2010.

\bibitem{Milligan2011}
M.~B. Milligan, {\em The E and B EXperiment: Implementation and Analysis of the
  2009 Engineering Flight}.
\newblock PhD thesis, University of Minnesota, 2011.

\bibitem{Sagiv2011}
I.~S. {Sagiv}, {\em {The EBEX Cryostat and Supporting Electronics}}.
\newblock PhD thesis, University of Minnesota, 2011.

\bibitem{Irwin2005}
K.~Irwin and G.~Hilton, ``Transition-edge sensors,'' in {\em Cryogenic Particle
  Detection},  C.~Enss, ed., {\em Topics in Applied Physics} {\bf 99},
  pp.~63--150, Springer Berlin Heidelberg, 2005.

\bibitem{Westbrook2012}
B.~Westbrook, A.~Lee, X.~Meng, A.~Suzuki, K.~Arnold, E.~Shirokoff, E.~George,
  F.~Aubin, M.~Dobbs, K.~MacDermid, {\em et~al.}, ``Design evolution of the
  spiderweb tes bolometer for cosmology applications,'' {\em Journal of Low
  Temperature Physics}~{\bf 167}(5-6), pp.~885--891, 2012.

\bibitem{Smecher2010}
G.~Smecher, F.~Aubin, E.~Bissonnette, M.~Dobbs, P.~Hyland, and K.~MacDermid,
  ``A biasing and demodulation system for kilopixel tes bolometer arrays,''
  {\em Instrumentation and Measurement, IEEE Transactions on}~{\bf 61},
  pp.~251--260, Jan 2012.

\bibitem{Dobbs2008}
M.~Dobbs, E.~Bissonnette, and H.~Spieler, ``Digital frequency domain
  multiplexer for millimeter-wavelength telescopes,'' {\em Nuclear Science,
  IEEE Transactions on}~{\bf 55}, pp.~21--26, Feb 2008.

\bibitem{sqlite}
``Sqlite.'' \url{http://www.sqlite.org}.

\bibitem{MacDermid2014}
K.~D. MacDermid, {\em Development and Performance of the Detectors and Readout
  of the EBEX Balloon-Borne CMB Polarimeter}.
\newblock PhD thesis, McGill University, 2014.

\bibitem{Mather1982}
J.~C. {Mather}, ``{Bolometer noise: nonequilibrium thoery},'' {\em Applied
  Optics}~{\bf 21}, pp.~1125--1129, Mar. 1982.

\end{thebibliography}
\bibliographystyle{spiebib}   

\end{document}